\documentclass[12pt,prd,tightenlines,nofootinbib,showpacs,showkeys]{revtex4}
\newcommand{\be}{\begin{equation}}
\newcommand{\ee}{\end{equation}}
\newcommand{\bga}{\begin{equation}\begin{gathered}}
\newcommand{\ega}{\end{gathered}\end{equation}}
\newcommand{\Tr}{\mathrm{Tr}}
\usepackage{graphics}
\usepackage{rotating}
\usepackage{epsfig}
\usepackage{amsmath}
\usepackage{amsfonts}

\begin{document}
\title{\begin{flushright}{\rm\normalsize SSU-HEP-13/02\\[5mm]}\end{flushright}
Relativistic corrections to $\eta_c$-pair production \\ in high energy
proton-proton collisions}
\author{\firstname{A.P.} \surname{Martynenko}}
\email{a.p.martynenko@samsu.ru}
\affiliation{Samara State University, Pavlov Street 1, 443011, Samara, Russia}
\affiliation{Samara State Aerospace University named after S.P. Korolyov, Moskovskoye Shosse 34,
443086, Samara, Russia}
\author{\firstname{A.M.} \surname{Trunin}}
\email{amtrnn@gmail.com}
\affiliation{Samara State Aerospace University named after S.P. Korolyov, Moskovskoye Shosse 34,
443086, Samara, Russia}

\begin{abstract}
On the basis of perturbative QCD and the relativistic quark model we calculate relativistic
corrections to the double $\eta_c$ meson production in proton-proton interactions at LHC energies.
Relativistic terms in the production amplitude connected with the relative
motion of heavy quarks and the transformation law of the bound state wave functions to the
reference frame of moving charmonia are taken into account. For the gluon and quark
propagators entering the amplitude we use a truncated expansion in relative quark momenta
up to the second order. Relativistic corrections to the quark bound state wave functions
are considered by means of the Breit-like potential. It turns out that the examined
effects decrease total nonrelativistic cross section more than two times and on 20
percents in the rapidity region of LHCb detector.
\end{abstract}

\pacs{13.85.Ni, 12.39.Ki, 12.38.Bx}

\keywords{Hadron production in proton-proton interaction, Relativistic quark model}

\maketitle

The high energy and luminosity of the LHC makes it possible to observe the pair charmonium production
processes and to measure the corresponding cross sections with sufficiently high accuracy.
The result of the measurement of the pair $J/\psi$ meson production by the LHCb Collaboration was published
in~\cite{LHCb} and discussed many times on different workshops \cite{belyaev}.
This process together with the charmonium and associated open charm
production can be considered as a probe of the quarkonium production mechanism.
According to non-relativistic QCD (NRQCD)~\cite{BBL} and collinear parton model~\cite{braaten,brambilla2011,kramer}
the predictions of observed cross section in the leading order in $\alpha_s$ can be obtained by the
use of parton distribution functions and a set of local non-perturbative charmonium production color singlet
and color octet matrix elements~\cite{likhoded1988,berezhnoy,qiao,ko,li}.
In the proton-proton collisions, additional contributions from other mechanisms, such as the
double parton scattering (DPS) or the intrinsic charm content of the proton to the total cross section
are possible~\cite{baranov,baranov1,ms1995}. The processes of quarkonium production in proton-proton interaction
are generally described using the scale-dependent parton density functions. They are calculated as
functions of the Bjorken variable $x$ at some factorization scale within the approach
of the Dokshitzer-Gribov-Lipatov-Altarelli-Parisi~(DGLAP) evolution equations. However, the double
charmonium production in $pp$ collisions
at high energies can be sensitive to the details of the parton kinematics. Therefore, it is more appropriate
to use the parton distributions unintegrated over the transverse momentum $k_t$ in the framework of
the $k_t$-factorization~\cite{baranov}. There exists another source of theoretical uncertainty related
with the pair charmonium production which gives essential modification of the cross sections. It is
connected with the account of relative motion of heavy quarks forming the bound states. As was
shown in~\cite{mt2012} these relativistic corrections significantly change the cross section of the pair
charmonium production in $pp$ interaction obtained in non-relativistic approximation. The detailed
investigation of this relativistic mechanism for exclusive double charmonium production in $e^+e^-$
annihilation~\cite{bodwin,ebert1,ebert2} evidently shows that it is impossible to obtain the reliable
theoretical predictions for observed quantities without an account of relativistic corrections. Finally,
the next to leading order QCD corrections to the production amplitudes also should be taken into account.

The strategy of experimental investigations can be directed on the study of such physical reactions in
$pp$ collisions in which one of the described mechanisms of quarkonium production is dominant. Unfortunately,
as we know at present time all enumerated mechanisms have important effect in the pair charmonium production
and their contributions to the total cross section should be taken into account to obtain high accuracy
theoretical result.

In this work we continue the study of relativistic effects in the inclusive pair charmonium
production by considering the process $p+p\to\eta_c+\eta_c+X$. Our calculation of the production cross
section is performed on the basis of relativistic quark model used previously for the investigation
of relativistic corrections to the other reaction $p+p\to J/\psi+J/\psi+X$ in~\cite{mt2012}.
We work within the single-parton scattering (SPS) mechanism in which the basic contribution to
the charmonium production is determined by the gluon-gluon fusion.
The aim of the present study consists also in the analysis of some uncertainties
regarding the choice of parton distribution functions (PDF).
In spite of existing difficulties in the detecting of $\eta_c$ meson pairs
it is thought that in new run of the LHC this process will be studied more successfully.

The differential cross section $d\sigma$ for the inclusive double charmonium production in
proton-proton interaction can be presented in the form of the convolution of partonic cross
section $d\sigma[gg\to \eta_c \eta_c]$ with the parton distribution functions of the
initial protons~\cite{likhoded1988,kramer,braaten,brambilla2011}:
\be
\label{eq:cs-plus-x}
d\sigma[p+p\to \eta_c+ \eta_c+X]=\int dx_1 dx_2  f_{g/p}(x_1,\mu) f_{g/p}(x_2,\mu)
d\sigma[gg\to \eta_c \eta_c],
\ee
where $f_{g/p}(x,\mu)$ is the partonic distribution function for the gluon in the proton,
$x_{1,2}$ are the longitudinal momentum fractions of gluons.
The cross section formula~\eqref{eq:cs-plus-x} contains the factorization of the long distance
PDFs and the short distance gluon fusion cross section $d\sigma[gg\to \eta_c \eta_c]$ with
the factorization scale $\mu$. Neglecting the proton mass and taking
the c.m. reference frame of initial protons with the beam along the $z$-axis
we can present the gluon on mass-shell momenta $k_{1,2}=x_{1,2}\frac{\sqrt{S}}{2}(1,0,0,\pm 1)$.
$\sqrt{S}$ is the center-of-mass energy in proton-proton collision.

In the quasipotential approach the double charmonium production amplitude for the basic parton
subprocess $g+g\to \eta_c+\eta_c$ can be expressed as a convolution of a perturbative production
amplitude of two $c$-quark and $\bar c$-antiquark pairs $\mathcal T(p_1,p_2;q_1,q_2)$ and the
quasipotential wave functions of the final mesons $\Psi_{\eta_c}$~\cite{ebert1,ebert2}:
\be
\label{eq:m-gen}
{\mathcal M}[gg\to \eta_c \eta_c](k_1,k_2,P,Q)=\int \! \frac{d\mathbf p}{(2\pi)^3}
\int \! \frac{d\mathbf q}{(2\pi)^3} \, \bar\Psi(p,P) \bar\Psi(q,Q) \otimes \mathcal T(p_1,p_2;q_1,q_2),
\ee
where $p_1$ and $p_2$ are four-momenta of $c$-quark and $\bar c$-antiquark in the pair
forming the first $\eta_c$ particle, and $q_2$ and $q_1$ are the appropriate four-momenta for
quark and antiquark in the second meson $\eta_c$. They are defined in subsequent transformations
in terms of total momenta $P(Q)$ and relative momenta $p(q)$ as follows:
\be
\label{eq:p-dec}
p_{1,2}=\frac12 P \pm p,\quad (pP)=0; \qquad q_{1,2}=\frac12 Q \pm q,\quad (qQ)=0,
\ee
This expression
describes the symmetrical escape of the $c$-quark and $\bar c$-antiquark from
the mass shell.
In Eq.~\eqref{eq:m-gen} we integrate over the relative three-momenta of quarks and antiquarks
in the final state.
The systematic account of all terms depending on the relative quark momenta $p$ and $q$
in~\eqref{eq:cs-plus-x} is
important for increasing the accuracy of the calculation.
$p=L_P(0,\mathbf p)$ and $q=L_Q(0,\mathbf q)$
are the relative four-momenta obtained by the Lorentz transformation of four-vectors
$(0,\mathbf p)$ and $(0,\mathbf q)$ to the reference frames moving with the four-momenta $P$~and~$Q$.

\begin{figure}[t!]
\center\includegraphics[scale=0.75]{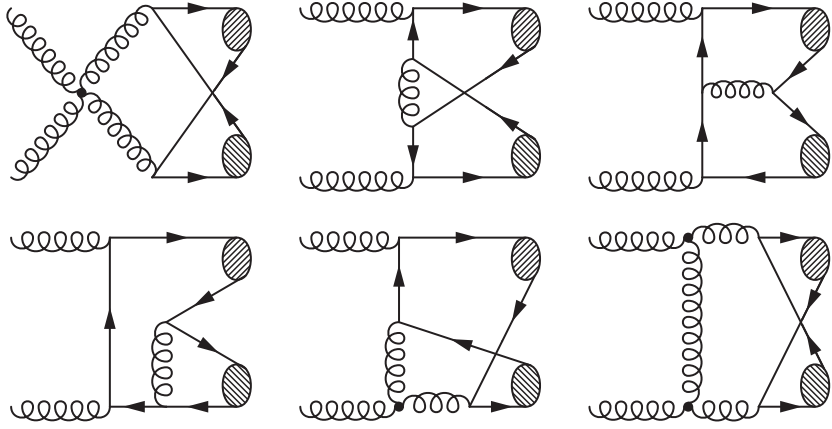}
\caption{The typical diagrams (the set of 31 Feynman diagrams) of the leading order
for $g+g\to\eta_c(J/\psi)+\eta_c(J/\psi)$.
The others can be obtained by reversing the quark lines or interchanging the initial gluons.
}
\label{fig:d31}
\end{figure}

The relativistic wave functions of the bound quarks, accounting for the transformation from the
rest frame to the moving one with four momenta $P$~and~$Q$, are the following~\cite{ebert1,ebert2}:
\bga
\label{eq:wf-transform}
\bar\Psi(p,P)=\frac{\bar\Psi_0^{\eta_c}(\mathbf p)}{\bigl[\frac{\epsilon(p)}{m}
\frac{\epsilon(p)+m}{2m}\bigr]}
\left[
\frac{\hat v_1-1}{2}+\hat v_1\frac{\mathbf p^2}{2m(\epsilon(p)+m)}-\frac{\hat p}{2m}
\right] \times \\
\gamma_5 \, (1+\hat v_1) \!
\left[
\frac{\hat v_1+1}{2}+\hat v_1\frac{\mathbf p^2}{2m(\epsilon(p)+m)}+\frac{\hat p}{2m}
\right],\\
\bar\Psi(q,Q)=\frac{\bar\Psi_0^{\eta_c}(\mathbf q)}{\bigl[\frac{\epsilon(q)}{m}
\frac{\epsilon(q)+m}{2m}\bigr]}
\left[\frac{\hat v_2-1}{2}+\hat v_2\frac{\mathbf q^2}{2m(\epsilon(q)+m)}+\frac{\hat q}{2m}
\right] \times \\
\gamma_5 \, (1+\hat v_2) \!
\left[
\frac{\hat v_2+1}{2}+\hat v_2\frac{\mathbf q^2}{2m(\epsilon(q)+m)}-\frac{\hat q}{2m}
\right],
\ega
where the hat symbol means a contraction of the four-vector with the Dirac gamma matrices;
$v_1=P/M$, $v_2=Q/M$; $\epsilon(p)=\sqrt{m^2+\mathbf p^2}$, $m$ is $c$-quark mass,
and $M$ is $\eta_c$ charmonium mass, $M\ne2m$.

The amplitude~\eqref{eq:m-gen} is projected onto a color singlet state by replacing
$v_i(0)\bar u_k(0)$ with a projection operator of the form
$v_i(0)\bar u_k(0)=\gamma_5(1+\gamma_0)\delta_{ik}/2\sqrt{6}$.
The relativistic wave functions in Eq.~\eqref{eq:wf-transform} are equal to the product
of the wave functions in the rest frame $\Psi_0^{\eta_c}$ and the spin projection operators that are
accurate at all orders in $|{\bf p}|/m$ \cite{ebert1,ebert2}. Our derivation of
relation~\eqref{eq:wf-transform} accounts for the transformation law of the bound state wave functions
from the rest frame to the moving one with four momenta $P$ and $Q$, which was obtained in
\cite{brodsky,faustov}. The physical
interpretation of the double charmonium production amplitude is the following:
we have a complicated transition of two heavy $c$-quark and $\bar c$-antiquark
produced in $gg$-fusion outside the mass shell and their
subsequent evolution firstly on the mass shell (free Dirac bispinors) and then to the
quark bound states. In the spin projectors~\eqref{eq:wf-transform} we have
${\bf p}^2$, ${\bf q}^2\not=M^2/4-m^2$ so that we can consider these structures as a transition
form factors for the heavy quarks from the free states to the bound states.

\begin{figure}[t!]
\center\includegraphics[scale=0.75]{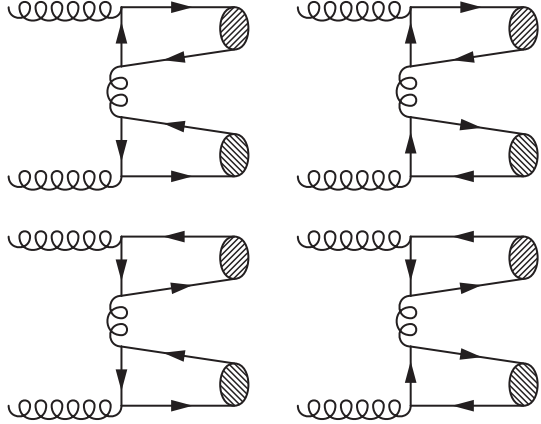}
\caption{The additional typical diagrams (the set of 8 Feynamn diagrams)
of the leading order for $g+g\to\eta_c+\eta_c$ only.}
\label{fig:d8}
\end{figure}

In the leading order in the strong coupling constant $\alpha_s$, there are 39 Feynman
diagrams contributing to the pair production of $\eta_c$ mesons. They can be divided into
two sets shown in Figs.~\ref{fig:d31} and \ref{fig:d8}, respectively. Their total contribution
to the production amplitude~\eqref{eq:m-gen} can be presented in the following form:
\bga
\label{eq:amp-LO}
{\mathcal M}[gg\to \eta_c\eta_c](k_1,k_2,P,Q)=
\frac19M\pi^2\alpha_s^2 \int\! \frac{d\mathbf p}{(2\pi)^3} \int\! \frac{d\mathbf q}{(2\pi)^3}
\bigl[\mathrm{Tr} \, \mathfrak M + 3\Delta\mathfrak M \bigr],
\ega
where we explicitly extracted the relativistic normalization factors
$\sqrt{2M}$ of quasipotential wave functions. The construction and transformation of the
production amplitudes is performed by means of the package FeynArts~\cite{feynarts}
for the system Mathematica and Form~\cite{form}.

The integrand term in~\eqref{eq:amp-LO} containing the trace of the amplitude $\mathfrak M$
represents the contribution of 31 diagrams in Fig.~\ref{fig:d31} and equals up to the wave functions
definitions~\eqref{eq:wf-transform} the analogous expression in the case of pair $J/\psi$
production, which can be found in Ref.~\cite{mt2012}. The second integrand term
in~\eqref{eq:amp-LO}, coming from additional 8 diagrams in Fig.~\ref{fig:d8} for $\eta_c$
production amplitude, has the form
\bga
\label{eq:amp-term2}
\Delta\mathfrak M=
\frac1t
\Tr\left[\hat\varepsilon_1\frac{m-\hat k_1+\hat p_1}{(k_1-p_1)^2-m^2}\gamma_\beta\bar\Psi(p,P)+
\gamma_\beta\frac{m+\hat k_1-\hat p_2}{(k_1-p_2)^2-m^2}\hat\varepsilon_1\bar\Psi(p,P) \right]\times \\
\Tr\left[\hat\varepsilon_2\frac{m-\hat k_2+\hat q_2}{(k_2-q_2)^2-m^2}\gamma^\beta\bar\Psi(q,Q)+
\gamma^\beta\frac{m+\hat k_2-\hat q_1}{(k_2-q_1)^2-m^2}\hat\varepsilon_2\bar\Psi(q,Q) \right]+
\\
\frac1{2M^2-s-t}
\Tr\left[\hat\varepsilon_1\frac{m-\hat k_1+\hat q_2}{(k_1-q_2)^2-m^2}\gamma_\beta\bar\Psi(q,Q)+
\gamma_\beta\frac{m+\hat k_1-\hat q_1}{(k_1-q_1)^2-m^2}\hat\varepsilon_1\bar\Psi(q,Q) \right]\times \\
\Tr\left[\hat\varepsilon_2\frac{m-\hat k_2+\hat p_1}{(k_2-p_1)^2-m^2}\gamma^\beta\bar\Psi(p,P)+
\gamma^\beta\frac{m+\hat k_2-\hat p_2}{(k_2-p_2)^2-m^2}\hat\varepsilon_2\bar\Psi(p,P) \right],
\ega
where the Mandelstam variables for the gluonic subprocess $gg\to\eta_c\eta_c$ are:
\begin{equation}
\label{eq:stu-def}
s=(k_1+k_2)^2=(P+Q)^2=x_1x_2S,
\end{equation}
\begin{equation}
t=(P-k_1)^2=(Q-k_2)^2=M^2-x_1\sqrt{S}(P_0-|{\bf P}|\cos\phi)=
\end{equation}
\begin{displaymath}
=M^2-x_1x_2S+x_2\sqrt{S}(P_0+|{\bf P}|\cos\phi),
\end{displaymath}
\begin{equation}
u=(P-k_2)^2=(Q-k_1)^2=M^2-x_2\sqrt{S}(P_0+|{\bf P}|\cos\phi)=
\end{equation}
\begin{displaymath}
=M^2-x_1x_2S+x_1\sqrt{S}(P_0-|{\bf P}|\cos\phi),
\end{displaymath}
$\phi$ is the angle between ${\bf P}$ and the $z$-axis.
The transverse momentum $P_T$ of $\eta_c$ and its energy $P_0$ can be written as
\be
\label{eq:pt0-def}
P_T^2=|{\bf P}|^2\sin^2\!\phi=-t-\frac{(M^2-t)^2}{x_1x_2S},\quad P_0=\frac{x_1x_2\sqrt{S}}{x_1+x_2}+
\frac{x_1-x_2}{x_1+x_2}|{\bf P}|\cos\phi.
\ee

In order to calculate relativistic corrections contained in the production
amplitude~\eqref{eq:amp-term2} we expand the inverse denominators of gluon and
quark propagators as series in relative quark momenta $p$ and $q$:
\bga
\label{eq:expansions}
\frac{1}{(p_1+q_1)^2}=\frac{4}{s}-\frac{16}{s^2} \left[ (p+q)^2+pQ+qP \right] +\cdots, \\
\frac{1}{(k_2-q_2)^2-m^2}=\frac{2}{t-M^2}-\frac{4}{\left(t-M^2\right)^2} \left[ q^2+2\,qk_2+\frac{M^2}{4}-m^2\right] +\cdots.
\ega
There are 16 different propagators in the amplitude~\eqref{eq:amp-LO}, which have to be expanded in the manner of
Eqs.~\eqref{eq:expansions}. Then, preserving in the expanded amplitude terms up to the
second order in relative quark momenta $p$ and $q$, we can perform angular
integration using the following relations for $\mathcal S$-wave charmonium:
\bga
\label{eq:wf-ints}
\int\!\frac{\Psi^{\mathcal S}_0(\mathbf p)}{\bigl[\frac{\epsilon(p)}{m}\frac{\epsilon(p)+m}{2m}\bigr]}
\frac{d\mathbf p}{(2\pi)^3}=\frac{1}{\sqrt{2}\,\pi}\int\limits_0^\infty\!\frac{p^2R_\mathcal S(p)}
{\bigl[\frac{\epsilon(p)}{m}\frac{\epsilon(p)+m}{2m}\bigr]}dp, \\
\int\! p_\mu p_\nu \, \frac{\Psi^{\mathcal S}_0(\mathbf p)}{\bigl[\frac{\epsilon(p)}{m}
\frac{\epsilon(p)+m}{2m}\bigr]}\frac{d\mathbf p}{(2\pi)^3}=-\frac{1}{3\sqrt2\,\pi}(g_{\mu\nu}-{v_1}_\mu{v_1}_\nu)
\int\limits_0^\infty\!\frac{p^4R_\mathcal S(p)}{\bigl[\frac{\epsilon(p)}{m}\frac{\epsilon(p)+m}{2m}\bigr]}dp,
\ega
where $R_\mathcal S(p)$ is the radial charmonium wave function.

As a result of the described transformations, we obtain the following general structure of pair
$\eta_c$ production amplitude~\eqref{eq:amp-LO}:
\bga
\label{eq:amp-final}
{\mathcal M}[gg\to \eta_c\eta_c]={\mathcal A}_1(\varepsilon_1\varepsilon_2)+
{\mathcal A}_2(\varepsilon_1P)(\varepsilon_2P)+{\mathcal A}_3(\varepsilon_1Q)(\varepsilon_2P)+\\
{\mathcal A}_4(\varepsilon_1P)(\varepsilon_2Q)+{\mathcal A}_5(\varepsilon_1Q)(\varepsilon_2Q),
\ega
where ${\mathcal A}_i$ are the functions of variables $s$ and $t$. Due to the bulkiness of
corresponding expressions for coefficient functions ${\mathcal A}_i$ we do not present them here
in exact form.

In order to find the differential cross section for the gluonic subprocess we should calculate
the squared modulus of the amplitude~\eqref{eq:amp-final} summed over polarizations of the
initial gluons by means of the following relation:
\bga
\label{eq:gluon-sum}
\sum_\lambda\varepsilon_{i}^\mu \, \varepsilon_{i}^{\ast\;\nu}=\frac{k_1^\mu k_2^\nu+k_1^\nu k_2^\mu}
{k_1k_2}-g^{\mu\nu}, \quad i=1,2.
\ega
Then we obtain the general form of the $gg\to\eta_c\eta_c$ cross section corresponding to the
production amplitude~\eqref{eq:amp-final}:
\bga
\label{eq:cs-pre}
\frac{d\sigma}{dt}[gg\to\eta_c\eta_c]=
\frac{1}{1024\,\pi s^4}\!\left(s^2{\mathcal A}_1^2+\left[s\mathcal A_1+(\mathcal A_2-
\mathcal A_3-\mathcal A_4+\mathcal A_5)\left(st+(M^2-t)^2\right)\right]^2\right).
\ega

Making the substitutions for the functions ${\mathcal A}_i$, we find it useful to transform the
result~\eqref{eq:cs-pre} as follows:
\bga
\label{eq:dcs}
\frac{d\sigma}{dt}[gg\to\eta_c\eta_c](s,t)=\frac{\pi\:\! M^2\alpha_s^4}{9216\,s^2}\,
|\tilde R(0)|^4\sum_{i=0}^3\omega_iF^{(i)}(s,t).
\ega
The auxiliary functions $F^{(i)}$ entering the cross section~\eqref{eq:dcs} are written
explicitly in Appendix. Note that the function $F^{(0)}$ describes non-relativistic result which
coincides in the limit $M_{\eta_c}=2m$ with the corresponding function obtained in Ref.~\cite{li}
in the approach of NRQCD. Relativistic corrections in~\eqref{eq:dcs} are determined by a number
of relativistic parameters $\omega_i$:
\bga
\label{eq:np-params}
\omega_0=1,\quad \omega_1=\frac{I_1}{I_0},\quad \omega_2=\frac{I_2}{I_0},\quad \omega_3=\omega_1^2,\\
I_0=\int\limits_0^\infty \frac{m+\epsilon(p)}{2\epsilon(p)}R(p) p^2 dp,\quad
I_{1,2}=\int\limits_0^m\frac{m+\epsilon(p)}{2\epsilon(p)}\left(\frac{m-\epsilon(p)}
{m+\epsilon(p)}\right)\!\negthickspace{\vphantom{\biggl|}}^{1,2} \! R(p) p^2 dp,\\
\tilde R(0)=\sqrt{\frac2\pi}\int\limits_0^\infty\! \frac{m+\epsilon(p)}{2\epsilon(p)}R(p)p^2 dp.
\ega
In the non-relativistic limit, the parameter $\tilde R(0)$ coincides with the definition of radial
wave function at the origin, so it can be considered in some way as its relativistic
generalization.

All parameters, which contain the meson wave functions and describe the transition of the pairs $(c\bar c)$
to the bound state, are calculated in the framework of relativistic quark model~\cite{ebert1,ebert2}.
This model is based on the Schr\"odinger equation with the Breit Hamiltonian in QCD and the
nonperturbative confinement terms. Using the program of numerical solution of the Schr\"odinger
equation~\cite{lucha1,mt2012}, we obtain relativistic wave functions and bound state energies of
$\mathcal S$-wave charmonia. Numerical values of charmonium masses $M_{J/\psi}^{th}=3.072$~GeV
and $M_{\eta_c}^{th}=2.988$~GeV obtained in our numerical calculation lie close to the experimental
results $M_{J/\psi}^{exp}=3.097$~GeV and $M_{\eta_c}^{exp}=2.981$~GeV~\cite{pdg2012}.
The additional details on our relativistic quark model can be found in Refs.~\cite{mt2012,ebert2}.

\begin{table}[h]
\caption{The comparison of relativistic and nonrelativistic cross sections of a pair $S$-wave
charmonium production in $pp$ collisions obtained for different sets of partonic distribution
functions.}
\bigskip
\label{tbl1}
\begin{tabular}{|l|l|c|c||c|c|} \hline
Energy $\sqrt{S}$ &
Meson pair, cross section type & \multicolumn{2}{|c||}{$\sigma$(total), nb} & \multicolumn{2}{|c|}
{$\sigma(2<y_{P,Q}<4.5)$, nb} \\  \cline{3-6}
  & & CTEQ5L & CTEQ6L1 & CTEQ5L & CTEQ6L1 \\   \hline
$\sqrt S=7$~TeV&
$J/\psi\, J/\psi$, relativistic & 9.6 & 7.4 & 1.6 & 1.2\\
&
$J/\psi\, J/\psi$, nonrelativistic & 23.0 & 17.7 & 3.8 & 2.9\\  \cline{2-6}
&
$\eta_c\, \eta_c$, relativistic & 23.7 & 19.9 & 1.3 & 1.0\\
&
$\eta_c\, \eta_c$, nonrelativistic & 56.3 & 48.1 & 1.5 & 1.2\\  \hline
$\sqrt S=14$~TeV&
$J/\psi\, J/\psi$, relativistic & 17.1 & 13.2 & 3.0 & 2.1\\
&
$J/\psi\, J/\psi$, nonrelativistic & 41.0 & 31.6 & 7.1 & 5.1\\  \cline{2-6}
&
$\eta_c\, \eta_c$, relativistic & 47.8 & 39.3 & 2.4 & 1.7\\
&
$\eta_c\, \eta_c$, nonrelativistic & 116.5 & 94.7 & 2.8 & 2.0\\  \hline
\end{tabular}
\end{table}

\begin{figure}[t]
\center\includegraphics[scale=0.9]{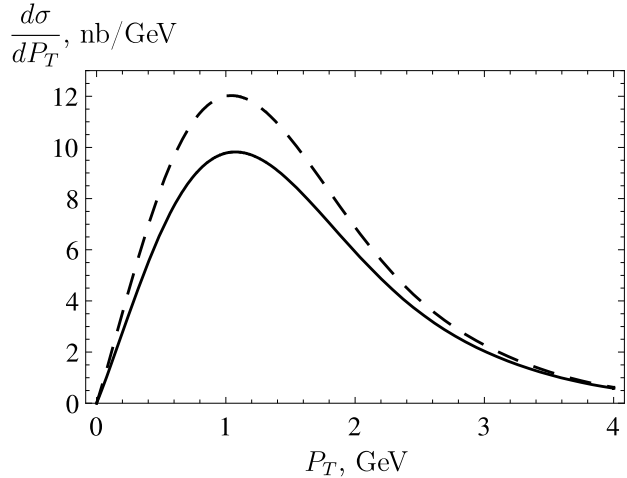}
\caption{The cross section $d\sigma/dP_T$ at $\sqrt{S}=7$~TeV for gluon distribution functions
CTEQ5L (dashed curve) and CTEQ6L1 (solid curve).}
\label{fig:pT}
\end{figure}

The numerical results of our calculation of the pair $\mathcal S$-wave charmonium production cross
sections in the case of non-relativistic approximation as well as with the account of relativistic corrections
of order $v^2$ are presented in Table~\ref{tbl1}. Along with total cross section values, we have also included there
the cross section predictions corresponding to the rapidity interval $2<y_{P,Q}<4.5$ of the
LHCb experiment~\cite{LHCb} calculated with two different sets of linear PDFs: CTEQ5L~\cite{CTEQ5} and
CTEQ6L1~\cite{CTEQ6}. As shown in Table I, the cross section~$\sigma[pp\to 2\eta_c+X]$ at $\sqrt{S}=7(14)$~TeV
is equal to 1.3 (2.4)~nb for CTEQ5L and 1.0 (1.7)~nb for CTEQ6L1. The most important production rates
lie in the region of small $P_T$ (see Fig.~\eqref{fig:pT}), where the color singlet contribution
is dominant. Performing the numerical integration of differential cross
section~\eqref{eq:dcs}, we use the LO expression for the running coupling constant $\alpha_s(\mu)$ with
the initial value $\alpha_s(M_Z)=0.118$ and the renormalization scale
$\mu=m_T=\sqrt{M^2+P_T^2}$, where $M$ is the meson mass. In our numerical calculations of the cross
sections we set $M_{\eta_c}=2.980$~GeV and $m=1.55$~GeV. Therefore, we take into account non-zero bound state
energy of $\eta_c$ charmonium state leading to the bound state corrections to the production cross
section~\eqref{eq:dcs}. Numerical results in Table I are determined by a number of parameters and functions:
the $c$-quark mass, the factorization scale $\mu$, parameters of the quark interaction operator,
the bound state wave functions, the parton distribution functions and the strong coupling constant.
Some of them (the $c$-quark mass,
the quark-antiquark potential) are fixed in the relativistic quark model in the mass spectrum
calculation. The factorization scale $\mu$ is taken in a commonly used form~\cite{berezhnoy,ko,li}.
Other quantities lead to basic uncertainties of our numerical results.

It is evident from Table~\ref{tbl1}, that relativistic corrections of order $v^2$ decrease the cross section
values more than two times in both cases connected with the pair production of $J/\psi$ or $\eta_c$
mesons. The only exception is the case of a pair $\eta_c$ production cross section in
the rapidity region $2<y_{P,Q}<4.5$, where the relativistic effects decrease the cross section
only by approximately 20 percents. The change of PDF from CTEQ5L to CTEQ6L1 brings the additional
$20\div 30$ percent decreasing to the value of cross section. Along with the possibility of different
PDF choices, there also exists the uncertainty dealt with the determination of every particular
partonic distribution function. The sets CTEQ5L and CTEQ6L1 contain no means to estimate the uncertainties
of such sort, however the set CTEQ6M~\cite{CTEQ6} has all necessary functionality. Using 40 uncertainty
eigenvectors from CTEQ6M we can roughly estimate the error of every cross section value in
Table~\ref{tbl1} dealt with the PDF uncertainty as 15~$\%$. The only known calculation of the
$\eta_c$-pair production in proton-proton collision was performed in Ref.~\cite{li}. Their Table~II
contains the obtained numerical results for different PDFs CTEQ5L and CTEQ6L1 with $P_T>3$~GeV, which are of
order of 4~nb. They used almost the same values of the $c$-quark mass, the factorization scale $\mu$
as in our calculation but a different numerical value for the parameter $R(0)$. Our value of the
radial wave function at the origin in nonrelativistic limit is equal to $R(0)\approx 0.8$~GeV$^{3/2}$, whereas
in Ref.~\cite{li} the authors took the long distance matrix element $\langle O_1\rangle_S=1.4$~GeV$^3$, which gave
$R(0)=1.7$~GeV$^{3/2}$. In the region of large transverse momentum
$P_T$ ($P_T>3$~GeV) the cross section falls considerably, so that the basic contribution to
our result in Table~I is determined by the region of small momenta $P_T$. Therefore, our
nonrelativistic results 1.5 nb and 1.2 nb differ significantly from the values of cross sections
obtained in nonrelativistic SPS approximation in~\cite{li}. This difference is related with
a choice of the parameter $R(0)$ in~\cite{li} which exceeds our value more than two times.

Another possible source of uncertainties is connected with the determination of relativistic
wave function in the momentum region $p\gtrsim m$. The obtained charmonium wave function
is strongly decreasing in this region.  Its numerical value at $p=m$ is more than 50 times smaller
the maximum value. Nevertheless, relativistic factors $p^2$ and $p^4$, entering in the integrals
$I_{1,2}$ change this relation and increase the inaccuracy in the wave function determination at
$p\gtrsim m$. In spite of the fact that momentum integrals appear to be
fully convergent, our relativistic model cannot provide a reliable calculation of the wave
functions in the region of relativistic momenta $p\gtrsim m$.
Our definitions~\eqref{eq:np-params} of the parameters $I_{1,2}$ contain the cutoff at relativistic
momentum of order~$m$. Using indirect arguments  related with the mass spectrum calculation accuracy
we estimate in $10\%$ the uncertainty of the wave function
determination. Larger value of the error would lead to the essential
discrepancy between the experiment and theory in the calculation of the charmonium mass
spectrum. Then the corresponding error in the cross section~\eqref{eq:dcs} is not
exceeding $20\%$. We do not consider a part of theoretical error related with
radiative corrections of order $\alpha_s$ because these corrections are omitted in our analysis.
We also assume that relativistic corrections of order $O(v^4)$ to the cross section~\eqref{eq:dcs}
coming from the production amplitude should not exceed $30\%$ of the obtained relativistic result.
So, our total theoretical error is not exceeding 39$\%$. To obtain this estimate we add the above
mentioned uncertainties in quadrature.

\acknowledgments

The authors are grateful to I.~Belyaev, A.V.~Berezhnoy, D.~Ebert, R.N.~Faustov, V.O.~Galkin for useful
discussions. The work is supported partially by the Ministry of Education and Science of
Russian Federation (government order for Samara State U. Grant No. 2.870.2011).

\appendix
\section{The coefficients $F^{(i)}$ entering the differential cross section~(\ref{eq:dcs})}

In this appendix, we present analytical results for the parton differential cross section~(\ref{eq:dcs}).
Firstly, we introduce the following auxiliary functions of the Mandelstam variables $s$, $t$, $u$,
and $\kappa=m/M$:

\bga
k_a=\frac{32}{3M^2s^3tu\bigl(2s+M^2(1-4\kappa^2)\bigr)\bigl(2t-M^2(1+4\kappa^2)
\bigr)^2\bigl(2u-M^2(1+4\kappa^2)\bigr)^2}\,,\\
k_b=\frac{-512}{9M^2s^3tu\bigl(2s+M^2(1-4\kappa^2)\bigr)^3\bigl(2t-M^2(1+4\kappa^2)\bigr)^4
\bigl(2u-M^2(1+4\kappa^2)\bigr)^4}\,,\\
k_c=\frac{256 (s - t + u)^2 (s + t - u)^2}{27M^2s^3tu(2s+M^2\bigl(1-4\kappa^2)\bigr)^3
\bigl(2u-M^2(1+4\kappa^2)\bigr)^6\bigl(2t-M^2(1+4\kappa^2)\bigr)^6}\,,
\ega

\be
\begin{split}
\label{eqa:1st}
&
a_1=3 s^8 (t+u)+148 s^7 t u-2 s^6 (t+u) (3 t^2-88 t u+3 u^2)-16 s^5 t (19 t^2-34 t u+19 u^2)u+\\&
s^4 (t-u)^2 (t+u) (3 t^2-446 t u+3 u^2)+4 s^3 t (t-5 u) (t-u)^2 (5 t-u) u+212 s^2 t (t-u)^4 \times\\&
(t+u) u +72 s t (t-u)^6 u+(\kappa-1/2)\bigl(9 s^8 (t + u) + s^7 (21 t^2 + 838 t u + 21 u^2) +2 s^6 \times\\&
(t + u) (3 t^2 + 790 t u + 3 u^2) -2 s^5 (9 t^4 + 54 t^3 u - 1142 t^2 u^2 + 54 t u^3 + 9 u^4) -5 s^4 \times\\&
(t + u) (3 t^4 + 340 t^3 u - 494 t^2 u^2 + 340 t u^3 + 3 u^4) -s^3 (t - u)^2 (3 t^4 + 980 t^3 u + 626 t^2 u^2 + \\&
980 t u^3 + 3 u^4) -16 s^2 t (t - u)^2 (17 t^2 - 86 t u + 17 u^2) (t + u) u -4 s t (t - u)^4 \times\\&
(55 t^2 + 38 t u + 55 u^2) u -72 t (t - u)^6 (t + u) u\bigr),
\end{split}
\ee

\be
\begin{split}
&
\Delta a=-16 s^2 t u \bigl(12 s^5 + 14 s^4 (t + u) -  s^3 (30 t^2 - 56 t u + 30 u^2) - 27 s^2 (t - u)^2 (t + u) + \\&
4 s (t - 2 u) (t - u)^2 (2 t - u) + 9 (t - u)^4 (t + u)\bigr)-8stu(\kappa-1/2)\bigl(143 s^6 + 245 s^5 (t + u) - \\&
12 s^4 (5 t^2 - 34 t u + 5 u^2) -2 s^3 (t + u) (109 t^2 - 158 t u + 109 u^2) -s^2(t - u)^2 \times \\&
(17 t^2 + 86 t u + 17 u^2) +s (t - u)^2 (t + u) (25 t^2 + 54 t u + 25 u^2) -14 (t - u)^4 (t + u)^2\bigr),
\end{split}
\ee

\be
\begin{split}
&
b_1=3 s^{14}(t + u)+2380 s^{13}t u-12s^{12}(t + u)(t^2-383 t u + u^2)-4 s^{11}t u\times\\&
(1471 t^2 - 5862 t u + 1471 u^2) +2 s^{10} (t + u) (9 t^4 - 8074 t^3 u + 18050 t^2 u^2 - 8074 t u^3 + 9 u^4) + \\&
4 s^9 t (t - u)^2 (65 t^2 - 10314 t u + 65 u^2) u -4 s^8 (t - u)^2 (t + u) (3 t^4 - 4479 t^3 u + 11992 t^2 u^2 - \\&
4479 t u^3 + 3 u^4) +20 s^7 t (t - u)^4 (413 t^2 + 1518 t u + 413 u^2) u +s^6 (t - u)^4 (t + u) \times\\&
(3 t^4 - 5312 t^3 u + 22906 t^2 u^2 - 5312 t u^3 + 3 u^4) -32 s^5 t (t - u)^6 (184 t^2 + 251 t u + 184 u^2) u - \\&
8 s^4 t (t - u)^6 (t + u) (259 t^2 - 22 t u + 259 u^2) u +8 s^3 t (t - u)^8(73 t^2 + 102 t u + 73 u^2) u + \\&
1008 s^2 t (t - u)^{10} (t + u) u + 288 s t (t - u)^{12} u-(\kappa-1/2)\bigl(9 s^{14} (t+u)-s^{13} \times\\&
(39 t^2+33074 t u+39 u^2)-4 s^{12} (t+u) (21 t^2+21470 t u+21 u^2)+4 s^{11} (27 t^4+2386 t^3 u- \\&
85150 t^2 u^2+2386 t u^3+27 u^4)+2s^{10} (t+u) (99 t^4+99076 t^3 u-270430 t^2 u^2+\\&
99076 t u^3+99 u^4)-2 s^9 (45 t^6 - 72738 t^5 u - 101909 t^4 u^2 + 414228 t^3 u^3 - 101909 t^2 u^4 -\\&
72738 t u^5+45 u^6)-4 s^8 (t-u)^2 (t+u)(45 t^4+18070 t^3 u-122638 t^2 u^2+18070 t u^3+45 u^4)+\\&
4 s^7 (t-u)^2 (3 t^6-34040 t^5 u+26081 t^4 u^2+83112 t^3 u^3+26081 t^2 u^4-34040 t u^5+3 u^6)+ \\&
s^6 (t-u)^4 (t+u)(57t^4-70124 t^3 u-72506 t^2 u^2-70124 t u^3+57 u^4)+s^5 (t-u)^4 \times\\&
(9 t^6-2350 t^5 u-63273 t^4 u^2+2204 t^3 u^3-63273 t^2 u^4-2350 t u^5+9 u^6)+24 s^4 t (t-u)^6\times\\&
(t+u) (1163 t^2-1906 t u+1163 u^2)u+32 s^3 t (t-u)^6 (562 t^4-577 t^3 u-18 t^2 u^2-\\&
577 t u^3+562u^4)u+40 s^2 t(t-u)^8 (t+u) (133 t^2-34 t u+133 u^2)u+144 s t (t-u)^{10} \times\\&
(19 t^2+46 t u+19u^2)u+864 t(t-u)^{12} (t+u)u\bigr),
\end{split}
\ee

\be
\begin{split}
&
\Delta b=-4 s^2 t u \bigl(1061 s^{11}+1859 s^{10} (t+u)-6 s^9 (433 t^2-1502 t u+433 u^2)-2 s^8 (t+u)\times\\&
(2827 t^2-6134 t u+2827 u^2)+24 s^7 (t-u)^2 (72 t^2-647 t u+72 u^2)+40 s^6 (t-u)^2 (t+u) \times\\&
(161 t^2-398 t u+161 u^2)+2 s^5 (t-u)^4 (313 t^2+5986 t u+313 u^2)-2 s^4 (t-u)^4 (t+u)\times\\&
(1525 t^2-4586 t u+1525 u^2)-s^3 (t-u)^6 (1189 t^2+3502 t u+1189 u^2)+s^2 (t-u)^6 (t+u)\times\\&
(261 t^2-1514 t u+261 u^2)+12 s (t-u)^8 (31 t^2+14 t u+31 u^2)+144 (t-u)^{10} (t+u)\bigr)-\\&
4stu(\kappa-1/2)\bigl(14421 s^{12}+34248 s^{11} (t+u)-s^{10} (8511 t^2-129586 t u+8511 u^2)- \\&
8 s^9 (t+u) (9129 t^2-23527 t u+9129 u^2)-2 s^8 (16065 t^4+59372 t^3 u-167130 t^2 u^2+\\&
59372 t u^3+16065 u^4)+4 s^7 (t-u)^2 (t+u) (10827 t^2-45008 t u+10827 u^2)+2 s^6 (t-u)^2 \times\\&
(18891 t^4+2840 t^3 u-77062 t^2 u^2+2840 t u^3+18891 u^4)+52 s^5 (t-u)^4 (t+u) \times\\&
(97 t^2+920 t u+97 u^2)-s^4 (t-u)^4 (6659 t^4-20876 t^3 u-3822 t^2 u^2-20876 t u^3+6659 u^4)- \\&
4 s^3 (t-u)^6 (t+u) (1869 t^2-2300 t u+1869 u^2)-s^2 (t-u)^6 (3191 t^4+1344 t^3 u-9454 t^2 u^2+\\&
1344 t u^3+3191 u^4)-4 s (t-u)^8 (t+u) (203 t^2+334 t u+203 u^2)-432 (t-u)^{10} (t+u)^2\bigr),
\end{split}
\ee
\be
\begin{split}
&
c_1=3 s^{14} (t+u)+127924 s^{13} t u-4 s^{12} (t+u) (3 t^2-99607 t u+3 u^2)-16 s^{11} t \times\\&
(6327 t^2-119146 t u+6327 u^2)u+2 s^{10} (t+u) (9 t^4-820440 t^3 u+1900702 t^2 u^2-\\&
820440 t u^3+9 u^4)-32 s^9 t (42787 t^4+95828 t^3 u-266062 t^2 u^2+95828 t u^3+42787 u^4)u-\\&
4 s^8 (t-u)^2 (t+u) (3 t^4-476492 t^3 u+1778482 t^2 u^2-476492 t u^3+3 u^4)+8 s^7 t (t-u)^2\times\\&
(395023 t^4+173924 t^3 u-1092582 t^2 u^2+173924 t u^3+395023 u^4)u+s^6 (t-u)^4 \times\\&
(t+u) (3 t^4-45892 t^3 u+6221826 t^2 u^2-45892 t u^3+3 u^4)-4 s^5 t (t-u)^4\times\\&
(587629 t^4-129076 t^3 u-999794 t^2 u^2-129076 t u^3+587629 u^4)u-4 s^4 t (t-u)^6 (t+u) \times\\&
(258537 t^2+588302 t u+258537 u^2)u+8 s^3 t (t-u)^6 (56911 t^4-55796 t^3 u-45238 t^2 u^2- \\&
55796 t u^3+56911 u^4)u+1440 s^2 t (t-u)^8 u (t+u) (287 t^2+166 t u+287 u^2)+\\&
288 s t (t-u)^{10} u (255 t^2+538 t u+255 u^2)-864 t (t-u)^{12} (t+u)u,
\end{split}
\ee

\be
\begin{split}
\label{eqa:6th}
&
\Delta c=-8t u(26215 s^{13}+72029 s^{12} (t+u)-s^{11} (46586 t^2-347812 t u+46586 u^2)-\\&
4 s^{10} (t+u) (78089 t^2-168598 t u+78089 u^2)-4 s^9 (37679 t^4+168700 t^3 u-401590 t^2 u^2+\\&
168700 t u^3+37679 u^4)+2 s^8 (t-u)^2 (t+u) (223529 t^2-615866 t u+223529 u^2)+\\&
2 s^7 (t-u)^2 (227477 t^4+195492 t^3 u-823282 t^2 u^2+195492 t u^3+227477 u^4)-\\&
4 s^6 (t-u)^4 (t+u) (50509 t^2-265278 t u+50509 u^2)-s^5 (t-u)^4 (402183 t^4-13884 t^3 u-\\&
817942 t^2 u^2-13884 t u^3+402183 u^4)-7 s^4 (t-u)^6 (t+u) (6701 t^2+56246 t u+6701 u^2)+\\&
24 s^3 (t-u)^6 (4913 t^4-2740 t^3 u-6138 t^2 u^2-2740 t u^3+4913 u^4)+48 s^2 (t-u)^8 (t+u) \times\\&
(817 t^2+812 t u+817 u^2)-324 s (t-u)^{10} (11 t^2+2 t u+11 u^2)-972 (t-u)^{12} (t+u)),
\end{split}
\ee

\bga
a_2=a_1+\Delta a,\quad b_2=b_1+\Delta b, \quad c_2=c_1+\Delta c.
\ega

Then, the coefficients $F^{(i)}$ entering the differential cross section~\eqref{eq:dcs} have
the following form:
\bga
F^{(0)}=k_a^2(a_1^2+a_2^2),\\
F^{(1)}=k_ak_b(a_1b_1+a_2b_2)+8k_a^2(a_1^2+a_2^2),\\
F^{(2)}=-4k_a^2(a_1^2+a_2^2),\\
F^{(3)}=6k_ak_b(a_1b_1+a_2b_2)+k_ak_c(a_1c_1+a_2c_2)+24k_a^2(a_1^2+a_2^2)+\frac14k_b^2(b_1^2+b_2^2).
\ega

Note that we expand functions $a_{1,2}$ and $b_{1,2}$ in the mass difference $(2m-M)$ up to
the term linear in $(\kappa-1/2)$ and set the value $\kappa=1/2$ in $c_{1,2}$. Such simplifications
allow us to significantly reduce the length of analytical expressions~\eqref{eqa:1st}--\eqref{eqa:6th},
while the numerical results of the cross sections change on $1\div 5$ percents. In Table~\ref{tbl1} we
present numerical results corresponding to exact functions $a_i$, $b_i$, and $c_i$.

\end{document}